# Field emission: calculations supporting a new methodology of comparing theory with experiment

Sergey V. Filippov[1], Anatoly G. Kolosko[1], Eugeni O. Popov[1] and Richard G. Forbes[2]

[1]Division of Plasma Physics, Atomic Physics and Astrophysics, Ioffe Institute, St. Petersburg 194021, Russia
[2]Advanced Technology Institute & Department of Electrical and Electronic Engineering, University of Surrey, Guildford, Surrey GU2 7XH, United Kingdom

E-mail: r.forbes@surrey.ac.uk

RGF, 0000-0002-8621-3298

**Abstract**

This paper presents new methodology for comparing field electron emission (FE) theory and experiment, and provides a "demonstration of concept". This methodology uses the exponent $\kappa$ that describes the power to which voltage is raised in the pre-exponential of a mathematical equation that describes (for an electronically ideal FE system) the dependence of measured current on measured voltage. It aims to use experimental exponent-values $\kappa^{\mathrm{expt}}$ to decide between two alternative FE theories, for which allowable (but different) $\kappa$-ranges have been established. At present, the contribution to the "total theoretical $\kappa$" made by the voltage-dependence of notional emission area is not well known: this paper report simulations that add relevant knowledge, for four commonly investigated emitter shapes. The methodology is then applied to choosing between 1928/29 Fowler-Nordheim (FN) FE theory and 1956 Murphy-Good (MG) FE theory. It is theoretically certain that 1956 MG FE theory is "better physics" than 1928/29 FN FE theory). Like all previous attempts to reach known correct theoretical conclusions by experimentally based argument, the new methodology tends to favour MG FE theory, but is formally indecisive at this stage. There seems an urgent need for better methods of measuring $\kappa^{\mathrm{expt}}$ and of establishing reliable error limits.

________________________________________________________________

## 1. Introduction

The work reported here contributes to a much wider project, recently outlined elsewhere [1] that aims to put field electron emission (FE) theory onto a more satisfactory scientific basis. As part of this wider project, it is wished to (a) improve existing methodologies for analysing FE current-voltage data and (b) improve existing procedures for using analysis results to derive useful scientific conclusions.

Sections 2 to 4 set out some detailed background to the present work; remaining Sections then present technical details and conclusions.

## 2. General background

The term *FE system* is defined to include all aspects of the experimental system that can affect the relationship between *measured current* $I_m$ and the *measured voltage* $V_m$, including: emitter composition, geometry and surface condition; the mechanical, geometrical and electrical arrangements in the vacuum system; all aspects of the electronic circuitry and all electronic measurement instruments; the emission physics; and ALL relevant physical processes that might be taking place (for example, the generation of field emitted vacuum space-charge, Maxwell-stress-induced reversible changes in emitter geometry, and adsorbate atom dynamics).

This $I_m(V_m)$ relationship can be discussed as follows. It is assumed that there exists a definite (albeit unknown) relationship $I_m(F_C)$ between $I_m$ and the magnitude $F_C$ of the electrostatic field at some defined *characteristic location* "C" on the emitter surface, and that the relationship between $F_C$ and $V_m$ can be written

$$F_C = V_m/\zeta_C, \qquad (1)$$

where $\zeta_C$ is a parameter called the (characteristic) *voltage conversion length (VCL)*. The VCL $\zeta_C$ is a system characterization parameter, not a physical length.

If the characteristic VCL is constant for a given FE system, and if there are no significant current- or time-dependent changes in the chemical state of the emitting surface (and hence no change in work-function values), then the FE system is said to be *electronically ideal*.

If the characteristic VCL is not constant, due to some "system complication" (e.g., significant series resistance, but many other possible complications exist), the FE system is *electronically non-ideal*. For such systems, the interpretation of measured FE current-voltage characteristics can be extremely complex, and often impossible to carry out exactly in the present state of research knowledge.

There currently exist two forms of *validity check* (namely, apparent linearity of a Fowler-Nordheim or similar data-analysis plot, and the Orthodoxy Test [2]), that can be used to assess whether it is likely that measured $I_m(V_m)$ data have been taken from an electronically ideal FE system.

For *electronically ideal* FE systems, the interpretation problem becomes part of FE emission physics, with the value of the characteristic VCL determined *in principle* by the zero-current electrostatics of the system geometry (but in practice nearly always by experiment). This paper is about FE current-voltage analysis in electronically ideal systems.

For the last 90 years, since the publication in 1928 [3] of the ground-breaking (but seriously flawed [4, 5]) paper of Fowler and Nordheim, most FE theory has been developed in the context of so-called *smooth planar metal-like emitter (SPME) methodology*. This methodology disregards the existence of atomic structure, uses Sommerfeld-type free-electron theory, and treats emitters as having smooth, structureless planar surfaces of large lateral extent. However, it is possible to treat needle-shaped emitters that are "not too sharply pointed" by using the *planar emission approximation* [1] and carrying out an integration of local emission current density (LECD) over the emitter surface.

As is well known, the main theoretical LECD equations in SPME methodology have been: (a) the original 1928 FN FE equation (as corrected in 1929 [6]), which used an exactly triangular (ET) transmission barrier; and (b) the 1956 Murphy-Good (MG) FE equation [7], which is based on the Schottky-Nordheim (SN) (or "planar image rounded") transmission barrier.

The most common method of analysing $I_m(V_m)$ data has been the well-known "Fowler-Nordheim (FN) plot" introduced by Stern et al. in 1929 [6]. The 1956 Murphy-Good FE theory is better physics than the 1928/29 Fowler-Nordheim FE theory (see [5] for a discussion) and predicts that a FN plot will be slightly curved. This curvature makes it difficult to reach accurate quantitative conclusions from the plot intercept on the vertical axis. However, a different plot form, the so-called Murphy-Good plot (see [8] for details) should (in SPME methodology) be "very nearly straight".

However, even for electronically ideal FE systems, it is not expected that an experimental MG plot should be exactly straight. This is because other factors, neglected in MG FE theory (and more generally in SPME methodology) are known to contribute additional voltage-dependence, and thus to cause departure from strict MG plot linearity. The main factors of this type are: (i) the long-known fact [9] that voltage dependence in notional emission area (see below) will be introduced when an experimental or theoretical integration of LECD over the needle surface is carried out; and (ii) the need to take atomic structure into account in a complete FE theory.

There are significant fundamental physical difficulties involved in *accurately* incorporating atomic structure into FE theory, and this has persuaded the authors to take a strategically different approach to making comparisons between FE theory and experiment. Rather than formulating some specific advanced FE theory, and then attempting to test this against experiment, there is a proposal [10,11] that attempts should be made to fit experimental data for FE systems proven to be electronically ideal to the *mathematical form*

$$I_m = CV_m^\kappa \exp[-B/V_m], \qquad (2)$$

where $B$, $C$ and $\kappa$ (also written "k" in some past work) are parameters that in the first instance have been treated as constants. Two methods of carrying out this fitting have been discussed in [11].

Our present expectation is that $C$ and $\kappa$ will be weakly varying functions of field, and that $B$ will be a constant or a very weakly varying function of field. Thus, depending on the data-analysis method used, the extracted values of $C$ and $\kappa$ may be average values for the range of voltages (and hence the range of characteristic-field values) used.

We emphasize that eq. (2) is a mathematical form, *not* a theory of FE. In particular, it should *not* be assumed that $\exp[-B/V_m]$ is an expression for the Gamow factor (or "barrier strength") that is used in FE theory. We also emphasize that eq. (2) is a mathematical form that is being applied in this paper to experimental data from FE systems assumed to be *electronically ideal* (i.e., systems for which the VCL is constant), *not* to FE $I_m(V_m)$ data in general—for which the VCL may be a function of measured voltage.

Equation (2) has previously been called the "empirical FE equation", but this name now seems potentially misleading. The main previous users of this mathematical form have been Abbott and Henderson in 1939 [9] (who used only integral values of $\kappa$), Forbes in 2008 [10], and Forbes, Popov, Kolosko and Filippov in 2019 [11]. Both [10] and [11] allowed $\kappa$ to be non-integral. Thus, this paper refers to eq. (2), with non-integral values allowed, as the *AHFP mathematical form* (for measured voltage).

The parameter "$\kappa$" has been called the "pre-exponential voltage exponent" [11]. However, for an electronically ideal FE system, both the characteristic field $F_C$ and the related characteristic scaled field $f_C$ (defined below) are directly proportional to $V_m$. Thus, when AHFP-type equations are formulated in terms of these alternative variables, factors of $F_C^\kappa$ or $f_C^\kappa$ will appear in the equation. Thus, particularly when discussing electronically ideal FE systems, it is considered preferable to call $\kappa$ the *AHFP exponent.*

In principle, our long-term aim *in the context of emission theory* is as follows. After confirming that a FE system under investigation is electronically ideal, and then extracting experimentally derived values of $B$, $C$ and $\kappa$, our aim is to use these values to deduce more detailed information about the nature of FE theory. It is expected that the values of $C$ and $\kappa$ will be more useful than the value of $B$, but that trying to interpret extracted values of $C$ and $\kappa$ simultaneously would be a difficult endeavour. Current efforts have therefore concentrated on issues related to the extraction and interpretation of values of $\kappa$ alone.

It has been shown by Forbes et al. [11], by simulations, that the best average value of $\kappa$ found (for an electronically ideal system) from a plot of $\ln\{I_m/V_m^\kappa\}$ vs $1/V_m$ (a so-called *power-$\kappa$ plot*) is a sensitive function of both (a) the details of emission theory and (b) the shape of the field emitter.

Hence, if $I_m(V_m)$ measurements of high quality were available for *electronically ideal* FE systems, and the effects of emitter shape and surface condition could be disentangled from these, then a measured value of $\kappa$ should provide a method of comparing different FE theories with experiment. Here we report some steps towards at least assessing emitter-shape effects.

An experimental alternative would be to make reliable direct measurements of local emission current density, but this seems to be significantly more difficult experimentally.

As part of the overall process, we need to know: (a) what physical effects contribute to the experimental value of $\kappa$; (b) what is the size of each contribution (or what are the limits within which each contribution size is likely to lie); and (c) what information is already available in the literature. Existing knowledge about these things is reviewed in Section 3, where it will become apparent that a gap in our knowledge is good understanding of how the voltage dependence of the notional emission area depends on the emitter shape and on assumed work-function value. An aim of this paper is to report the result of simulations relating to this knowledge gap. This is done in Sections 3 and 4. However, we need first to provide, in Section 3.1 some additional details of existing theory.

## 3. Emission-theory background

*3.1 Summary of Extended Murphy-Good (EMG) FE current-voltage theory*

For our simulations, we use the *Extended Murphy-Good (EMG)* formulation (see [8]) of FE theory. The local *kernel current density for the SN barrier* ($J_{kL}^{SN}$) is given by:

$$J_{kL}^{SN} = a\phi^{-1}F_L^2 \exp[-v_F b\phi^{3/2}/F_L], \qquad (3)$$

where the symbols have their usual meanings [12]. This equation can be put into *scaled format* by defining three parameters:

$$\eta(\phi) \equiv bc_S^2\phi^{-1/2}, \qquad (4)$$

$$\theta(\phi) \equiv a\, c_S^{-4}\phi^3, \qquad (5)$$

$$f_L \equiv c_S^2\, \phi^{-2} F_L, \qquad (6)$$

where $c_S\,[\equiv (e^3/4\pi\varepsilon_0)^{1/2}]$ is the *Schottky constant*, and $f_L$ is the local *scaled-field* for a barrier of zero-field height $\phi$. Using these yields the *scaled-format expression*:

$$J_{kL}^{SN} = \theta f_L^2 \exp[-v(f_L)\cdot\eta/f_L], \qquad (7)$$

In EMG theory, the *local emission current density (LECD)* $J_L^{EMG}$ is written

$$J_L^{EMG} = \lambda_L^{SN} J_{kL}^{SN}, \qquad (8)$$

where the parameter $\lambda_L^{SN}$ is a *prediction uncertainty factor* of unknown value and functional form. It formally takes account of all relevant physics not included in 1956 MG FE theory, including atomic-level wave-function effects and band-structure effects, and also incorporates the usual Murphy-Good (MG) pre-exponential correction factor $t_F^{-2}$, and the MG temperature correction factor.

If the kernel current density $J_{kL}^{SN}$ is integrated over the surface of a needle-shaped emitter, using the planar emission approximation, then the resulting calculated *notional emission current* $I_n^{SN}$ can be written in the form

$$I_n^{SN} = A_{nC}^{SN} J_{kC}^{SN}, \tag{9}$$

where $J_{kC}^{SN}$ is the value of the kernel current density at some surface location "C" that characterizes the emitter. (In modelling, "C" is usually taken as the emitter apex). The *notional emission area* $A_{nC}^{SN}$ is defined by eq. (9); it depends on the chosen emitter shape and work-function distribution, on the assumed nature of the barrier (here the SN barrier), and on the choice of location "C".

We may suppose that, for the chosen emitter model, there would be a true (but unknown) model emission current $I_{tm}$, and that we can write;

$$I_{tm} = \lambda_J^{SN} I_n^{SN} \tag{10},$$

where $\lambda_J^{SN}$ is a *prediction uncertainty factor for current density*, of the same general kind as $\lambda_L^{SN}$. The factor $\lambda_J^{SN}$ will depend on many things, but we specifically show the barrier label: this is because the interpretation of experimental results (in particular, FN and related plots) requires an assumption about the nature of the transmission barrier.

In reality, when attempting to accurately model the "predicted" emission current $I_p$ from a real emitter, it is inevitable (at least at present) that the model emitter will not be an accurate model for the real emitter. Hence a second source of prediction uncertainty is introduced, and we can write

$$I_p = \lambda_{EM}^{SN} I_{tm} = \lambda_{EM}^{SN} \lambda_J^{SN} I_n^{SN}, \tag{11}$$

where $\lambda_{EM}^{SN}$ *is a prediction uncertainty factor related to emitter-model inadequacy.*

Using eq. (9) yields

$$I_p = \lambda_{EM}^{SN} \lambda_J^{SN} I_n^{SN} = \lambda_{EM}^{SN} \lambda_J^{SN} A_{nC}^{SN} J_{kC}^{SN} \tag{12}$$

This can be simplified by defining a new area-like parameter $A_{fC}^{SN}$, *the formal emission area for the SN barrier* (as defined by location "C") by

$$A_{fC}^{SN} = \lambda_{EM}^{SN} \lambda_J^{SN} A_{nC}^{SN}. \tag{13}$$

Equation (12) can then be re-written as

$$I_p = A_{fC}^{SN} J_{kC}^{SN}. \tag{14}$$

Leakage current, if present, can be modelled by an additional correction factor included within $\lambda_{EM}^{SN}$.

We now assert that (for the electronically ideal systems under discussion) the "formally predicted" emission current $I_p$ can be identified with the measured current $I_m$, and hence [using eq. (7)] we can write

$$I_m = A_{fC}^{SN} J_{kC}^{SN} = A_{fC}^{SN} \theta f_C^2 \exp[-v(f_C) \cdot \eta/f_C], \tag{15}$$

where $f_C$ is the scaled-field at location "C".

In eq. (15), $I_m$ is well defined, and $J_{kC}^{SN}$ is well defined in principle (if location "C" is known, and the values of $\phi$ and $f_C$ are stated). Hence, the formal area $A_{fC}^{SN}$ is well defined in principle. It is this *formal* emission area that is extracted from experiments, typically by using FN plots.

For an electronically ideal FE system, the parameter $f_C$ can also be written $f_C = V_m/V_{mR}$, where $V_{mR}$ is the *reference measured voltage* needed to pull the top of the SN barrier down to the Fermi level, at location "C".

Our assumption is that, in a first approximation (at least over defined ranges of measured voltage $V_m$), $A_{nC}^{SN}$ can be written in the form

$$A_{nC}^{SN} = C_A V_m^{\kappa_A} = \{C_A V_{mR}^{\kappa_A}\} f_C^{\kappa_A}, \tag{16}$$

where $C_A$ and $\kappa_A$ are parameters associated with the shape and work-function characteristics of the emitter. As already indicated, we expect these parameters either to be constants, or (more likely) to be weakly varying functions of apex field and hence of measured voltage.

The discussion above is given specifically for the SN barrier and EMG theory, but generalized versions of the theory can be given for other barrier forms.

*3.2 Individual contributions*

For clarity, in this Section the individual contributions to the total $\kappa$-value are denoted by the symbol $\Delta\kappa$. The factors currently known to affect the experimental $\kappa$–value are as follows.

*A. Emitter band-structure.* A three-dimensional free-electron model of emitter band structure, as used in deriving the 1928/29 FN and 1956 MG FE equations, makes a contribution $\Delta\kappa=2$. Models involving lower dimensionality will make, and situations involving quantum confinement may make, a contribution less than 2. It is not clearly known what the effects of more-complicated real band-structures, such as the "many-valley" band-structure of silicon [13] are expected to be.

*B. Barrier form (i.e., shape).* As has been shown in Refs. [10, 11, 14, 15], there exist formulations of FE theory in which a SN barrier makes a contribution to $\kappa$. (For an emitter with local work function 4.50 eV, $\Delta\kappa \approx -0.773$). Barriers similar in form to the SN barrier may also behave in this way.

*C. Atomic-level structure.* In Oppenheimer's approach [16, 17], FE is interpreted as the field ionization (FI) of the atomic-level orbitals of surface metal atoms. The existing standard theory of the FI of a hydrogen atom [18] is equivalent to setting $\Delta\kappa = -1$. It might reasonably be assumed that a FE theory that in effect involved FI out of surface metal atoms might make a contribution somewhere in the range $-1 \leq \Delta\kappa < 0$. Some relevant work (e.g., [16–20]) currently exists, but greater depth is probably needed.

*D. Voltage dependence in the notional emission area.* As discussed above, a formula for *notional* emission current $I_n$ will involve a notional emission area $A_n$, and this will usually have field and voltage dependence. Details will depend on the chosen emitter shape and the work-function distribution. There is some relevant work, e.g. [12, 14, 15, 21, 22], but no good general knowledge about the range of $\Delta\kappa$-values involved. The present work aims to help fill this gap.

*E Other physical factors included in the prediction uncertainty factor ($\lambda_J$) related to emission theory.* As indicated above, correction factors related to other physical effects (for example, a temperature correction factor) are formally included in $\lambda_J$. These will or may contribute to $\kappa$. Our present understanding is that—certainly for bulk metals with conventional 3-D band-structures—contributions due to these "other factors" are likely to be relatively small and hence can be disregarded for the time being. It is currently not clear whether this is also likely for materials with non-conventional band-structures.

*F. The prediction uncertainty factor ($\lambda_{EM}$) related to emitter-model inadequacy.* When predicting emission currents, if an inadequate emitter model is used, then it is likely that an incorrect total $\kappa$-value will be predicted, and that a correction will be necessary. At present, there is no useful classified knowledge about $\lambda_{EM}$ or this correction, and this paper does not discuss this issue in detail.

Table 1 provides a summary of existing knowledge relating to the AHFP pre-exponential exponent $\kappa$. Experimental data has been included only if the related FE experimental system is known or thought to be electronically ideal.

Table 1. Research results relating to the AHFP exponent $\kappa$ for an electronically ideal FE system. For theoretical sources, the labels "A" (etc.) in the "category" column indicate which of the individual contributions noted in the text were taken into account.

| Row | Year | Ref. | Structure | Category | Total $\kappa$-value |
|---|---|---|---|---|---|
| 1 | 1928 | Oppenheimer [16, 17] | Surface metal atom | Theor (C) | 0.25 |
| 2 | 1928 | Millikan & Lauritsen [23], Lauritsen [24] | Thoriated tungsten cylinder | Expt | 0 |
| 3 | 1928 | Fowler & Nordheim [3] | SPME[a] | Theor (A) | 2.00 |
| 4 | 1929 | Stern et al. [6] [re-analysis of [25]][b] | Thoriated tungsten cylinder | Expt | 2 |
| 5 | 1929 | Stern et al. [6] | Pointed tungsten wire | Expt | 2 |
| 6 | 1939 | Abbott & Henderson [9] | Pointed tungsten wire[c] | Expt | 4 |
| 7 | 1939 | Abbott & Henderson [9] | SPME + PEA[a] | Theor (A+D) | 3 |
| 8 | 1958 | Landau & Lifschitz [18] | Hydrogen atom | Theor (C) | −1.00 |
| 9 | 2008 | Forbes [10] | SPME ($\phi$=4.50 eV) | Theor (A+B) | 1.23 |
| 10 | 2014 | Jensen [14,15] | SPME+PEA, hemisphere-on-plane | Theor (A+B+D) | 1.99 to 2.13 |
| 11 | 2019 | Popov et al. [26] | CNT LAFE[d] | Expt | 1.65 |

| 12 | 2020 | Ang et al. [27][e] | 2D-Dirac/Weyl semimetals | Theor (A+E) | 1.00 |
| 13 | 2021 | Lepetit [28] | Graphene | Theor[f] ~(A+B+C+E) | 1.53 |
| 14 | 2021 | Chan et al. [29] | 3D-Dirac/Weyl semimetals | Theor (A+E) | 3.00 |
| 15 | 2022 | Biswas [30][g] | SPME+PEA Single tip "with generic endform" | Theor (A+B+D) | 3–$\eta$/6 [≈2.23][h] |

[a]SPME = "smooth planar metal-like emitter"; PEA = "planar emission approximation".
[c]Re-analysis of experimental results obtained by Millikan and Eyring [25].
[c]In their "precise measurements", Abbott and Henderson used "point-plane" geometry,
[d]CNT = "carbon nanotube"; LAFE = "large area field electron emitter".
[e]Also see: https://arxiv.org/abs/2003.14004 .
[f]Lepetit's advanced theoretical treatment does not easily fit into the classification system used above.
[g]The argument is also made by Biswas that the behaviour of LAFEs is much more complicated than that of single-tip emitters, and involves a contribution to the main exponent.
[h]The value 2.23 applies to an emitting surface with local work function 4.50 eV.

A further account of how notional area depends on apex field has recently been presented by Ramachandran and Biswas [22]. This is discussed in Section 6.

### *3.3 The concept of notional cap-area efficiency*

As already indicated, it is of considerable theoretical interest to know how the notional emission area $A_{nC}$ varies as a function of characteristic field $F_C$ and how this functionality depends on the emitter shape and work-function characteristics. In practice, we use cylindrically symmetric models, and take "C" as located at the emitter model apex "a". Also, it is more useful to investigate the behaviour of the so-called *notional cap-area efficiency* $g_n$ (also called the "area factor") given by

$$g_n = A_{na} / 2\pi r_a^2, \qquad (17)$$

where $r_a$ is the *apex radius of curvature* of the model. Further, it is better to investigate $g_n$ as a function of the *apex scaled-field* $f_a$.

Results will depend on what assumption is made as to the form of the transmission barrier. All discussion here assumes a Schottky-Nordheim barrier, and that the planar emission approximation is being used. [Except that to compile Table 3 below, additional calculations have been carried out, using an exactly triangular barrier.]

For the special case of a hemisphere on a plane (HSP case), Jensen ([14], see eq. (94), or [15] see eq. (30.19)] has obtained an analytical approximation, which he writes in the form "$g(F) \approx 1/[b+4-v]$". In our notation this becomes

$$g_n(f_a) \approx 1/[\eta/f_a + 4 - \eta/6], \qquad \text{(HSP)} \qquad (18)$$

$$1/g_\mathrm{n} \approx (4 - \eta/6) + \eta/f_\mathrm{a} . \qquad \text{(HSP)} \qquad (19)$$

However, what we actually need is the form of the relationship $\ln\{g_\mathrm{n}\}$ vs $\ln\{f_\mathrm{a}\}$, so that we can establish the power-law dependence of $g_\mathrm{n}$ on $f_\mathrm{a}$ and hence (for electronically ideal systems) on $V_\mathrm{m}$. Straightforward algebraic manipulation leads to the result

$$\Delta\kappa = \mathrm{d}\ln\{g_\mathrm{n}\}/\mathrm{d}\ln\{f_\mathrm{a}\} \approx 1/[1+ \{(4/\eta)-(1/6)\}f_\mathrm{a}] . \qquad (20)$$

For $\phi = 4.50$ eV, $\eta = 4.6368$, and eq. (19) yields

$$\Delta\kappa \approx 1/[1+ 0.696\,f_\mathrm{a}] , \qquad \text{(HSP)} \qquad (21)$$

$$1/\Delta\kappa \approx 1+ 0.696\,f_\mathrm{a} . \qquad \text{(HSP)} \qquad (22)$$

So, over the range $0.15 \le f_\mathrm{a} \le 0.45$ (which is the "Pass" range for the orthodoxy test [2]), $\Delta\kappa$ takes values in the range $0.905 \ge \Delta\kappa \ge 0.761$, for the HSP analytical model.

Jensen has also given a formula ([14], see eq. (113)) for the case of a hemi-ellipsoid on a plane. He writes his formula as "$g(F) \approx 1/[b+1-v]$", which in our notation becomes

$$g_\mathrm{n}(f_\mathrm{a}) \approx 1/[\eta/f_\mathrm{a}+1- \eta/6], \qquad \text{(HEP)}. \qquad (23)$$

However, there seems to be some difficulty with this formula, because it does not reduce to the HSP formula when the aspect ratio (height/base-radius) goes to unity.

Since these are approximate analytical results, we have thought it helpful to check their accuracy by means of finite-element electrostatic simulations, as discussed below. We then went on to carry out numerical analyses for emitter shapes of greater practical interest, for which analytical treatment looks either impossible or impracticable.

.

## 4. Simulation details

In this and following Sections, for historical reasons, the contribution (to the total value of $\kappa$) made by field/voltage dependence in the notional emission area is denoted by $\kappa_A$, rather than by $\Delta\kappa$.

### 4.1 Procedures

To solve the Laplace equation numerically, a commercial finite-element software package (COMSOL v5.3) was used. In addition to the hemisphere-on-plane (HSP) model, we studied three well-known emitter shape models, namely the hemisphere-on-cylindrical post (HCP), the spherically-rounded cone (SRC), and the hemi-ellipsoid-on plane (HEP). Recently [31], it has been shown that the apex field enhancement factors for the hemi-ellipsoidal, paraboloidal and hyperboloidal (with half-angle 5º) posts are similar, so we do not consider paraboloidal and hyperboloidal posts here. Details of our simulation procedure are similar to those found in Refs. [11,31].

The emitters (other than the HSP model) are given the following geometric parameters: total height $h= 4$ μm, and apex radius of curvature $r_\mathrm{a}= 50$ nm. This implies that the *apex sharpness ratio* $\sigma_\mathrm{a}$

= $h/r_a$ = 80. The SRC emitter has an additional parameter $\theta$, which defines the half-angle of the vertex (here $\theta$ = 5º). The work function of the emitters was taken as consta set to 4.50 eV.

A cylindrical simulation cell was used. The cell radius and height, the meshing geometry, and the electrostatic boundary conditions, were chosen in accordance with the "minimum domain size", meshing criteria and electrostatic arguments outlined in [32]. As an illustration, Figure 1a shows the resulting field distribution for a particular case of the hemisphere-on plane (HSP) model.

Integration of current density over the emitter surface, to evaluate the emission current, was carried out directly in the software package using the planar emission approximation and the kernel current density for the Schottky-Nordheim barrier.

*4.2 Consistency check with Jensen's analytical treatment*

For an initial comparison between our present results for the HSP model and those of Jensen, we show (in Fig. 1(a)) our results in the form as used by Jensen in Fig. 30.3 of Ref. [15], namely a direct plot of $g_n$ versus the apex field-magnitude $F_a$. (But note that Jensen uses a quantity measured in eV/nm to represent the behaviour of electrostatic fields: his quantity "$F$" equals our "$eF$"). This is done for the three work-function values Jensen used. Our plot and Jensen's plot are closely similar.

For $\phi$ = 4.50 eV, Fig. 1(b) shows a more detailed comparison. The *percentage relative difference $\delta$* between the analytical and simulation results is

$$\delta \equiv 100\ \% \times |\ g_n^{\text{simulation}} - g_n^{\text{analytical}}\ | / g_n^{\text{analytical}}. \tag{24}$$

The inset to Fig. 1(c) shows how $\delta$ varies with $f_a$: $\delta$ decreases almost linearly as the apex field magnitude $F_a$ (and hence the scaled field $f_a$) increase. Over the "Pass" range of the Orthodoxy Test ($0.15 \leq f_a \leq 0.45$) the percentage relative difference lies approximately in the range ($-3\% \leq \delta \leq -1\%$).

Obviously, there is a question of whether this small difference is due to "approximation error" in the analytical calculations or to "simulation error". However, we know from the work of de Assis and Dall'Agnol [32], on minimum simulation domain dimensions, that the simulations ought to be highly precise. Thus, our expectation is that most of the difference $\delta$ arises from the mathematical approximations made in the analytical treatment.

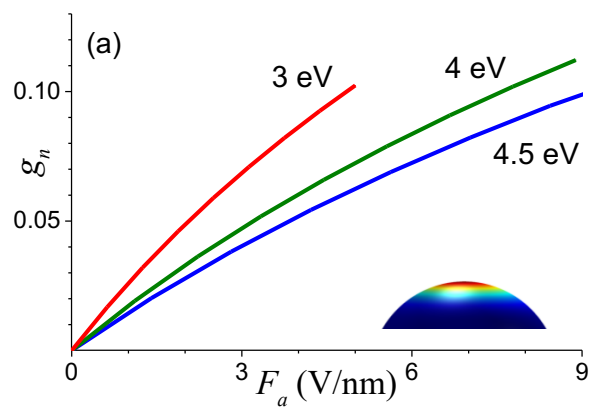

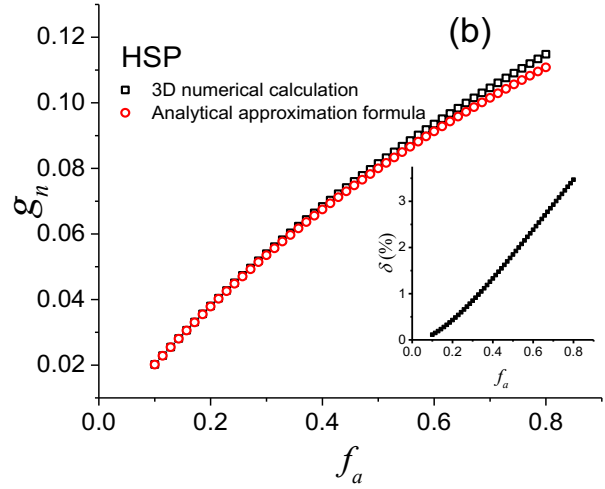

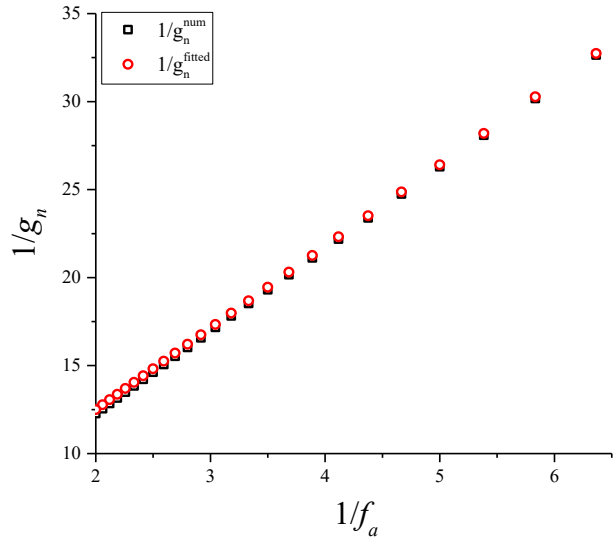

Fig.1. Hemisphere-on-plane (HSP) emitter model. (a) Dependence of notional cap-area efficiency $g_n$ on local apex-field $F_a$, for three different values shown of local work function. (Inset – 3D finite-element simulation of the electrostatic field for HSP). (b) Comparison (for a $\phi$=4.50 eV emitter) of numerically calculated values of notional cap-area efficiency and values of Jensen's analytical formula (inset – relative percentage difference $\delta$.) (c) To show (for a $\phi$=4.50 eV emitter) that $1/g_n$ is a nearly linear function of $1/f_a$ : the black squares are the finite-element numerical results and the red circles are the corresponding values derived from eq. (18).

For $\phi$= 4.50 eV, Fig. 1(c) confirms the $1/g_n$ is a good linear function of $1/f_a$ .

A further comparison can be made between the plots of $\kappa_A(f_a)$ derived from eq. (21) and from our finite-element calculations. This is shown in Fig. 2. The degree of agreement is very satisfactory. More generally, Fig. 2 clearly predicts that $\kappa_A$ will be a noticeable function of $f_a$ (and hence of apex field and of measured voltage). For the circumstances of these calculations, in the range $0.15 \leq f_a \leq 0.45$, $\kappa_A$ takes values (for the SN barrier) in the approximate range $0.9 \geq \kappa_A \geq 0.75$.

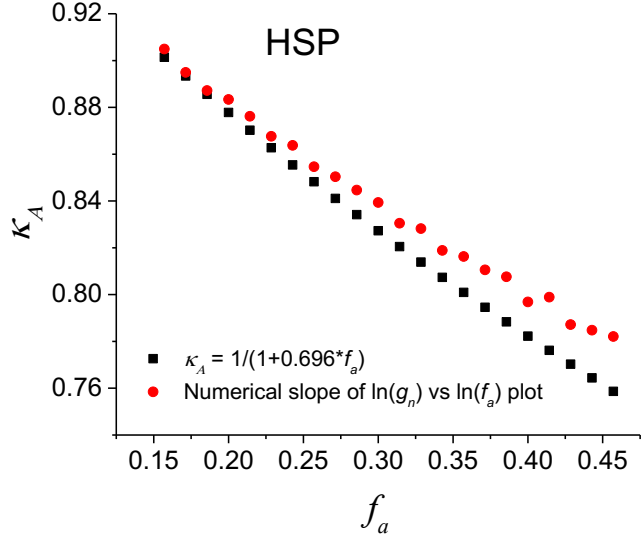

Fig. 2. For the HSP emitter model, using the kernel current density for the SN barrier for simulations, with $\phi = 4.50$ eV, Fig. 2 shows the contribution $\kappa_A$ (to the AHFP exponent) resulting from field dependence in the calculated notional emission area. The figure compares the results of our finite-element electrostatic simulations (red circles) with the predictions of eq. (16) (black squares).

## 5. Simulation results and discussion

The hemisphere-on-plane (HSP) model is the easiest to use when making comparisons between analytical models and simulations, and provides a useful degree of verification for the simulations. However, the HSP model is obviously not a realistic field-emitter model. Other field emitter shapes, closer to experimental reality, are of greater interest. As indicated earlier, we have investigated three other emitter models: the hemisphere-on-cylindrical-post (HCP) model, the hemiellipsoid-on-plane (HEP), and the spherically rounded (truncated) cone (SRC) model. For completeness, the data for the hemisphere-on-plane (HSP) model discussed in Section 4 are also included in diagrams below.

The current-voltage characteristics were calculated using the general procedure described in Section 4.1. As before, an emitter height of 4 μm and apex radius of 50 nm (corresponding to an apex sharpness ratio of 80) were used, and calculations were carried out for apex scaled-fields in the range $0.15 \leq f_a \leq 0.45$.

Figure 3 shows the results of the current density integration over the emitter surface, but with the current plotted as a function of the macroscopic field $F_M$ that is applied between the top and bottom surfaces of the simulation box. The corresponding apex field enhancement factors ($\gamma_a$) are also indicated.

As illustrated by the graphs, the lowest threshold values of macroscopic field (and hence of applied voltage) correspond to sharply pointed elongated shapes, as is well known. The shapes of individual carbon nanotubes correspond most closely to the HCP model, which is the model most effective at field enhancement. This is part of the reason why carbon nanotubes have been extensively investigated as field emitters in recent years.

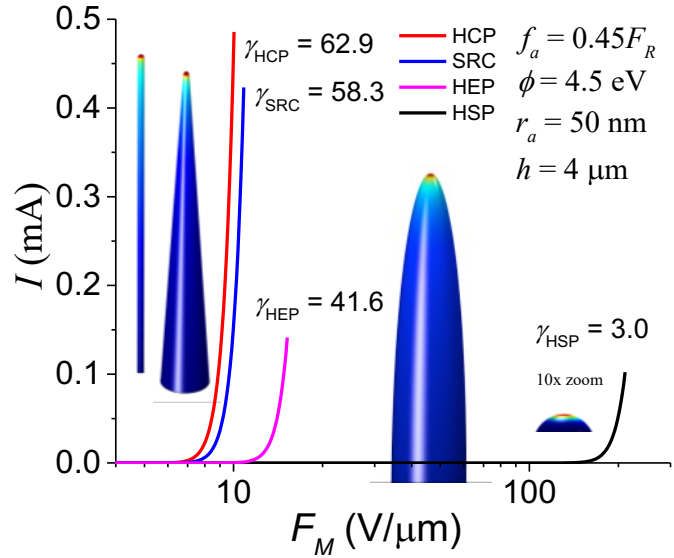

Fig. 3. Data for the HSP, HEP, SRC and HCP models, for $\phi$=4.50 and an apex-sharpness ratio (for the latter three models) of 80. The figure shows shapes, apex values of the field enhancement factor ($\gamma$), illustrative surface field distributions, and plots of total emission currents $I$ as functions of the macroscopic field $F_M$ applied between the top and bottom surfaces of the simulation box, over macroscopic field ranges corresponding to apex scaled-field values up to $f_a$=0.45.

Figure 4 shows how the notional cap-area efficiency $g_n$ varies as a function of apex scaled field $f_a$, for the four emitter shapes investigated. Clearly, these results fall into two groups. It might be expected intuitively that emitters with similar apex shapes would exhibit similar behaviour. However, the highest $g_n$ values were in fact obtained for the HCP model.

This result was not obviously expected. However, closer examination shows that the HCP emitter has not only has the largest apex FEF, but also has higher field values along a line along the surface, in a plane containing the model axis. This leads to greater integral current values, and, therefore, to larger values of notional emission area (and, hence to larger values of $g_n$).

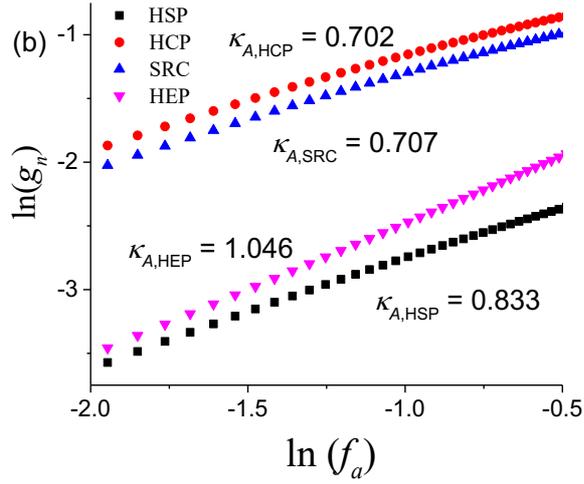

Fig. 4. Dependences of the notional cap-area efficiency $g_n$ on the (dimensionless) apex scaled field $f_a$, in the range $0.15 \leq f_a \leq 0.45$, for the four model emitters investigated, as specified in Fig. 3. Estimates of effective values of $\kappa_A$, derived from the mean slopes of these functions over the stated range, are shown.

To investigate numerically how the quantity $\kappa_A$ (the contribution to the AHFP exponent due to emitter-shape effects) varies with $f_a$ (and hence, for electronically ideal emitters, with measured voltage), plots of $d\ln\{g_n\}/d\ln\{f_a\}$ vs $f_a$ were made for the studied tip shapes. These plots are shown in Fig. 5.

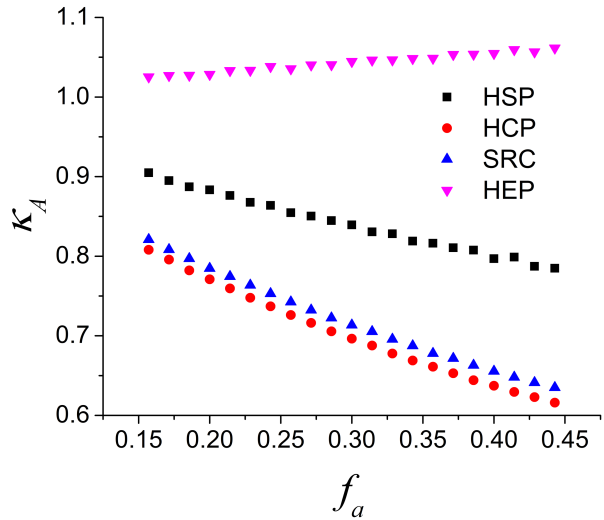

Figure 5. The contribution $\kappa_A$ (to the AHFP exponent) resulting from field dependence in the calculated notional emission area, for each of the emitter models investigated, as a function of the (dimensionless) apex scaled-field $f_a$.

Figure 5 suggests that, for all tip shapes, there will be a contribution $\kappa_A$ (to the total AHFP exponent) that is a function of the apex scaled-field $f_a$, and hence of the independent variable in an AHFP-type equation. For the range $0.15 \leq f_a \leq 0.45$ under discussion, Table 2 records the variation in $\kappa_A$, and also its average value $(\kappa_A)_{av}$ over the range (obtained by linear regression to the data plots in Fig. 5).

Table 2. To show, for the apex-scaled-field range $0.15 \leq f_a \leq 0.45$, and the four emitter shapes investigated (assuming a SN barrier), the limits of the variation of the contribution $\kappa_A(f_a)$ across the range, and its average value $(\kappa_A)_{av}$. Combining the individual four ranges provides an estimate of the total likely range of variation.

| Shape | $\kappa_A(f_a)$ | | $(\kappa_A)_{av}$ |
|---|---|---|---|
| | $f_a$=0.15 | $f_a$=0.45 | |
| HSP | 0.90 | 0.78 | 0.833 |
| HEP | 1.02 | 1.06 | 1.046 |
| HCP | 0.81 | 0.62 | 0.702 |
| SRC(5°) | 0.82 | 0.63 | 0.707 |
| *Total range* | 1.02 | 0.62 | n/a |

## 6. A methodology for using $\kappa$-values in theory-experiment comparisons

With the theory and simulations now described, we now present a new methodology for comparing FE theory and experiment. The methodology seeks to use an experimental $\kappa$–value to decide (if possible) which of two specific FE theories is "better science". As an example, we shall seek to establish *experimentally* that the Schottky-Nordheim barrier used by Murphy and Good in 1956 is "better science" than the exactly triangular barrier used by Fowler and Nordheim in 1928. Theoretically, there is no doubt about this (see [5]), but establishing a *decisive* proof from experiment has proved elusive (e.g., see [33]).

We emphasise that the remarks here aim to establish a methodology, and provide a "proof of concept", but do not aim to produce a definitive result at this stage. We anticipate that getting a decisive result may be a long and messy process, but we hope that establishing this methodology may stimulate the development of techniques able to provide good experimental values of $\kappa$, with reliable error limits.

For the 1928 Fowler-Nordheim FE equation (as corrected in 1929) and the 1956 Murphy-Good FE equation, both as expressed in AHFP mathematical form, Table 3 collates the various individual contributions (as defined in Section 3.1) to the total AHFP exponent $\kappa$. Entries on line F could arise if an experimental emitter of interest were not adequately modelled by any of the emitter shapes we have examined (or not adequately modelled by the assumption of constant local work-function), but at present we have no good knowledge of effects of this kind.

| Table 3. Range limits on contributions to the AHFP exponent and its total value, for current-voltage characteristics taken from an electronically ideal, needle-shaped or post-shaped emitter, according to (a) 1928/29 Fowler Nordheim FE theory and (b) 1956 Murphy-Good FE theory, for an emitter with uniform local work function 4.50 eV. (Contribution sources are defined in Section 3.1 .) | | | | |
|---|---|---|---|---|
| Contributions: sources of $\Delta\kappa$: | (a) FN (ET barrier) | | (b) MG (SN barrier) | |
| | lower limit | upper limit | lower limit | upper limit |
| A: | 2 | 2 | 2 | 2 |
| B: | 0 | 0 | −0.77 | −0.77 |
| C: | −1 | 0 | −1 | 0 |
| D: | 0.58 | 1.03 | 0.62 | 1.02 |
| E: | ~0 | ~0 | ~0 | ~0 |
| F: | undefined | Undefined | undefined | undefined |
| Range of total $\kappa$: | 1.58 | 3.03 | 0.85 | 2.25 |

In each case, on the last line of Table 3, we arrive at an estimate of the range within which $\kappa$ is predicted to lie if the equation in question is an adequate representation of experimental reality. These two ranges can then be used to generate Table 4, which shows how an experimental estimate $\kappa^{\text{expt}}$ of the (total) AHFP exponent is to be interpreted, for emitters of the kind under discussion. It is stressed that the numbers in this table are "first estimates", based on our current theoretical assumptions, and may change as theoretical knowledge improves.

Table 4. Decision table for interpreting an experimental value $\kappa^{\mathrm{expt}}$ of the total AHFP exponent, for current-voltage characteristics taken from an electronically ideal needle-shaped emitter or a post-shaped emitter with apex sharpness ratio (height/apex -radius) 80, assuming a uniform local work function of 4.50 eV.

| $I(V)$-ideal: (needle/post) | Decision Ranges (as at 7Jun22) | Deduction |
|---|---|---|
| X: | $\kappa^{\mathrm{expt}} < 0.85$ | Not compatible with either theory |
| MG56: | $0.85 \leq \kappa^{\mathrm{expt}} < 1.58$ | 1956 MG theory indicated |
| ? | $1.58 \leq \kappa^{\mathrm{expt}} \leq 2.25$ | INDECISIVE: compatible with MG or with FN |
| FN29: | $2.25 < \kappa^{\mathrm{expt}} \leq 3.02$ | 1928/29 FN theory indicated |
| X: | $3.02 < \kappa^{\mathrm{expt}}$ | Not compatible with either theory |

It is of interest to apply this decision table to the small number of relevant experimental $\kappa^{\mathrm{expt}}$ values listed in Table 1.

The Lauritsen value of "0" (on row 2) was a highly important early empirical result, but has low accuracy: it also has restricted relevance to the present analysis, because the emission was coming from a small area (probably a pointed protrusion) on a cylindrical wire [24]. The re-analysis by Stern et al. (on row 4) of the Millikan and Eyring results [25] also has restricted relevance, for the same reason.

The analysis by Stern et al, of the Gossling results (on row 5) (Fig. 2 in their paper) had the objective of deciding whether $\kappa$=0 or $\kappa$=2 better represented experimental reality. Historically, the finding (on the basis of two Gossling data plots, and re-analysis of one Millikan and Eyring data plot) was that $\kappa$=2 was the better choice, in accordance with FN's theory. This finding led to the whole apparatus of Fowler-Nordheim plots, still in current use 90 years later.

However, if we apply the thinking of Table 4 to the Stern et al. conclusion that the Gossling data plots imply that $\kappa^{\mathrm{expt}}$=2, then this result of "2" is in the "*indecisive*" decision regime, thus apparently showing that the Gossling data plots are also compatible with 1956 MG FE theory. This amounts to a basic *experimentally-derived* prediction that FN plots are not necessarily expected to be exactly straight, which is in accordance with modern theoretical thinking. (Exact straightness of a FN plot is predicted by 1928/29 FN FE theory but not by 1956 MG FE theory.)

In 1939, the "precise measurements" by Abbott and Henderson reported in Part II of their paper led them to conclude that the best integral value was $\kappa^{\mathrm{expt}}$=4 (see row 6). According to Table 4, this result is not compatible with either FN FE theory or MG FE theory. This finding has never been satisfactorily explained and has usually been ignored. However, Abbott and Henderson were the first to point out (more than 80 years ago) that the value of $\kappa^{\text{expt}}$ must depend on the "exact microscopic shape of the point (emitter)"—as now well demonstrated by our present calculations.

Basic problems with all this early work are the limited accuracy of the experiments, and the lack of realization that $\kappa^{\text{expt}}$ could legitimately be non-integral (e.g., see [10]).

Turning to discuss modern work on $\kappa^{\text{expt}}$, the only table entry (on row 11) is to the Ioffe Institute work on carbon nanotubes. If we make the usual assumption that traditional metal emission theory is "good enough" for CNT data analysis, then applying Table 4 thinking to the row 11 result that $\kappa^{\text{expt}}$ =1.65 yields the result that these experiments are marginally indecisive.

In summary, the existing experimental evidence, as listed in Table 1, does not enable us to reach a decisive experimentally based decision that 1956 MG FE theory is a better description of experimental reality than 1928/29 FN FE theory, notwithstanding that it is theoretically certain that 1956 MG FE theory is "better physics" than 1928/29 FN FE theory, Thus, there is a strong need for appropriate and accurate new experiments, preferably on metal emitters.

## 6. Summary and discussion

The main focus of this research has been the prediction of values for the AHFP exponent $\kappa$, and on preliminary comparison of predicted values with such experimental results as currently exist. So-called Extended Murphy-Good field electron emission theory has been outlined, six main theoretical contributions to $\kappa$ (for current-voltage characteristics taken from electronically ideal FE systems) have been identified, and existing knowledge about these contributions (as far as we are aware of it) has been collated into Table 1. A gap in this knowledge, namely the detailed effect of emitter shape on the related contribution $\kappa_A$, has been filled (at least partially) by simulations on four different emitter tip shapes, for the specific common work-function value 4.50 eV. These shapes include the hemisphere-on-plane model, for which there is an analytical treatment by Jensen. For this model, our simulations and Jensen's analytics are in very good agreement.

A new methodology for using an experimental value $\kappa^{\text{expt}}$ of the AHFP exponent to decide which of two specific theories of electronically ideal FE current-voltage characteristics is "better science" has been outlined, and has been applied to the choice between the 1928/29 FN and 1956 MG FE theories. Based on current theoretical knowledge, the relevant decision table (Table 4) has been constructed, and applied both to historical values of $\kappa^{\text{expt}}$ and to the one modern value that we consider reliable.

As indicated earlier, an interesting result relates to the Stern et al. analysis of two data plots made in Gossling's laboratory at the General Electric Company in London in the late 1920s. Stern et al. concluded that taking $\kappa^{\text{expt}}=2$ gave lower residual error than taking $\kappa^{\text{expt}}=0$, and hence that the results were compatible with 1928/29 FN FE theory. As already indicated, the whole apparatus of Fowler-Nordheim plots, still in use 90 years later, was built on this finding (and on a related data plot generated by Millikan and Eyring). But the decision table in the present paper shows that the Gossling result of $\kappa^{\text{expt}}$=2 is also compatible with 1956 MG FE theory, and hence indicates *from experimental analysis* that FN plots are not the only suitable option for analysing FE current-voltage data. (An alternative is the so-called Murphy-Good plot [8], introduced on the basis of theoretical arguments.)

The single modern estimate of $\kappa^{\text{expt}}$ in the table tends to favour 1956 MG FE theory rather than 1928/29 FN FE theory, but is marginally indecisive. Thus, FE science still remains in the state in which it has been since the late 1950s: it is theoretically certain that MG FE theory is "better physics" than FN FE theory, but demonstrating this experimentally by a decisive argument remains elusive. The most important need looks to be for better experimental methodologies for accurate measurement of $\kappa^{\text{expt}}$ for metal emitters, and for determining reliable error limits.

We make again the general point that this paper represents "first steps" toward a new methodology. As theoretical knowledge improves, and experimental results get better defined and more accurate, the decision table will change numerically (and perhaps new contributions to $\kappa$ will be added), and it is expected that its interpretation will change qualitatively. When we have reached the point that this methodology can clearly show, in alignment with theoretical understanding, that 1956 MG FE theory is "better science" than 1928/29 FN FE theory, then it will be timely to construct one or two new decision tables that compare 1956 MG FE theory with one or two more-sophisticated FE theories, for example the 2014 Kyritsakis-Xanthakis theory [34] of emission from an earthed sphere.

We note that our simulations have so far been carried out only for one value of work function and one value of apex sharpness ratio. This has been sufficient to establish a proof of concept for our new methodology, but we also plan to carry out further simulations for ranges of values of these parameters.

Finally, two recent papers by other authors need comment. For a model involving a hemiellipsoid on a cylindrical post (HECP model), Ramachandran and Biswas [22] have investigated how the dependence of the notional emission area on the apex field varies with the ratio "hemi-ellipsoid height/apex radius". Detailed investigation of the relationship between their results and ours is outside the scope of this paper, but our intention is to incorporate an appropriately customized version of their results into updated versions of Tables 2 to 4, in due course.

In the second recent paper, Ayari et al. [35] have discussed whether methodology based on measuring $\kappa^{\text{expt}}$ could be used to extract a value for the local work-function $\phi$ from FE current-voltage measurements. We agree with their conclusion that this would be a very challenging task, but we think that the discussion in the present paper shows that the complete theoretical picture is more complex than their arguments have assumed (although the conclusion would be the same).

Finally, we state again our belief that the decision-table methodology introduced here could become a useful tool in developing better field electron emission science.

**Acknowledgements**

The authors are grateful to Prof. Fernando F. Dall'Agnol (Federal University of Santa Catarina) for his help in COMSOL simulations.